
\input harvmac
\noblackbox

\def\B{{\cal B}}
\def\K{{\cal K}}


%
%
\def\RF#1#2{\if*#1\ref#1{#2.}\else#1\fi}
\def\NRF#1#2{\if*#1\nref#1{#2.}\fi}
\def\refdef#1#2#3{\def#1{*}\def#2{#3}}
\def\rdef#1#2#3#4#5{\refdef#1#2{#3, `#4', #5}}

%
%
\def\ts{\hskip .16667em\relax}

\def\CMP{{\it Commun.\ts Math.\ts Phys.\ts}}

\def\FAP{{\it Funct.\ts Analy.\ts Appl.\ts}}

\def\JP{{\it J.\ts Phys.\ts}}

\def\NP{{\it Nucl.\ts Phys.\ts}}
\def\PL{{\it Phys.\ts Lett.\ts}}

\def\TMP{{\it Theor.\ts Math.\ts Phys.\ts}}

\def\Zm{Zamolodchikov}
\def\AZm{A.B. \Zm}
\def\dur{H.\ts W.\ts Braden, E.\ts Corrigan, P.E. Dorey \ and R.\ts Sasaki}
%
%
\rdef\rBCDRa\BCDRa{P. Bowcock, E. Corrigan, P.E. Dorey and R. H. Rietdijk}
{Classically integrable boundary conditions for affine Toda field theories}
{\NP {\bf B445} (1995) 469}

\rdef\rBCDSc\BCDSc{\dur}
{Affine Toda field theory and exact S-matrices}
{\NP {\bf B338} (1990) 689}

\rdef\rCo\Co{I.\ts V.\ts Cherednik}
{Factorizing particles on a half line and root systems}
{\TMP {\bf 61} (1984) 977}

\rdef\rCDRa\CDRa{E.\ts Corrigan, P.E. Dorey, R.H.\ts Rietdijk}
{Aspects of affine  Toda field theory on a half line}
{to appear in Proceedings of
\lq Quantum field theory, integrable models and beyond',
Yukawa Institute for Theoretical Physics, Kyoto University, February 1994;
hep-th/9407148}

\rdef\rCDRSa\CDRSa{E.\ts Corrigan, P.E. Dorey, R.H.\ts Rietdijk and R.\ts
Sasaki}
{Affine Toda field theory on a half line}
{\PL {\bf B333} (1994) 83}

\rdef\rFKc\FKc{A.\ts Fring and R.\ts K\"oberle}
{Factorized scattering in the presence of reflecting boundaries}
{\NP {\bf B421} (1994) 159}

\rdef\rFKd\FKd{A.\ts Fring and R.\ts K\"oberle}
{Affine Toda field theory in the presence of reflecting boundaries}
{\NP {\bf B419} (1994) 647}

\rdef\rFSa\FSa{A. Fujii and R. Sasaki}
{Boundary effects in integrable field theory on a half line}
{YITP-K-1091; hep-th/9503083}

\rdef\rHa\Ha{T.J. Hollowood}
{Solitons in affine Toda field theories}
{\NP {\bf 384} (1992) 523}

\rdef\rKa\Ka{J.D. Kim}
{Boundary reflection matrix in perturbative quantum field theory}
{\PL {\bf B353} (1995) 213}

\rdef\rKb\Kb{J.D. Kim}
{Boundary reflection matrix for A-D-E affine Toda field theory}
{DTP-95-31; hep-th/9506031}

\rdef\rMv\Mv{A. MacIntyre}
{Integrable boundary conditions for classical sine-Gordon theory}
{{\it  J. Phys.} {\bf A28} (1995) 1089}

\rdef\rGZa\GZa{S.\ts Ghoshal and \AZm}
{Boundary $S$ matrix and boundary state in two-dimensional
integrable quantum
field theory}
{{\it Int. J. Mod. Phys.} {\bf A9} (1994) 3841}

\rdef\rMOPa\MOPa{A. V. Mikhailov, M. A. Olshanetsky and A. M. Perelomov}
{Two-dimensional generalised Toda lattice}
{\CMP {\bf 79} (1981) 473}

\rdef\rOTa\OTa{D. I. Olive and N. Turok}
{The symmetries of Dynkin diagrams and the reduction of Toda field equations}
 {{\it Nucl. Phys.} {\bf B215} (1983) 470}

\rdef\rSk\Sk{R.\ts Sasaki}
{Reflection bootstrap equations for Toda field theory}
{in {\it Interface between Physics and Mathematics}, eds W. Nahm and J-M Shen,
(World Scientific 1994) 201}

\rdef\rSl\Sl{E.\ts K.\ts Sklyanin}
{Boundary conditions for integrable equations}
{\FAP {\bf 21} (1987) 164}

\rdef\rSm\Sm{E.\ts K.\ts Sklyanin}
{Boundary conditions for integrable quantum systems}
{\JP {\bf A21} (1988) 2375}

\rdef\rWa\Wa{G. Wilson}
{The modified Lax and and two-dimensional Toda lattice equations
associated with simple Lie algebras}
{{\it Ergod. Th. and Dynam. Sys.} {\bf 1} (1981) 361}

\rdef\rYa\Ya{A. Yegulalp}
{New boundary conformal field theories indexed by the simply-laced
Lie algebras}
{\NP {\bf B450} (1995) 641}

\rightline{DTP-95/41}
\rightline{hep-th/9510071}
\bigskip
\centerline{\bf Background field boundary conditions for affine Toda
field theories}
\vskip 3pc
\centerline{P. Bowcock, E. Corrigan, R.H. Rietdijk}
\bigskip
\centerline{Department of Mathematical Sciences}
\centerline{University of Durham}
\centerline{Durham DH1 3LE, England}
\medskip
\vskip 4pc
\centerline{\bf Abstract}
\bigskip
Classical integrability is investigated for affine Toda field theories
in the presence of a constant background tensor field. This leads to
a further set of discrete possibilities for integrable boundary
conditions depending on the time derivative of the fields at the
boundary but containing no free parameters
other than the bulk coupling constant.
\vskip 8pc
\noindent October 1995

\vfill
\eject

\newsec{Introduction}

There has been some interest  in analysing the
classical and quantum integrability of two-dimensional
field theories with boundary, in which a theory is
either restricted to a  half-line, or to an interval.
Some years ago, Cherednik
and Sklyanin
\NRF\rCo{\Co\semi\Sl\semi\Sm}\refs{\rCo}
developed a general mathematical machinery
which generalised the standard tools applicable to integrable
models to those cases in which a boundary condition needs to
be taken into account. Principally, these tools are generalisations of
the Yang-Baxter equations incorporating reflections from
the boundary: the so-called reflection equations, and their
classical counterparts.

Subsequently, Fring and K\"oberle,
\NRF\rFKc{\FKc\semi\FKd}
Ghoshal and Zamolodchikov,
\NRF\rGZa\GZa   Sasaki,
\NRF\rSk\Sk and Kim,
\NRF\rKa{\Ka\semi\Kb} \refs{\rFKc -\rKa}, have developed a set of
conjectures for the reflection factors of the
sine-Gordon and affine Toda field theory models, based on a
generalisation of the bootstrap idea. However, these conjectures
are not easily related to specified boundary conditions\foot{The
exception to this is Kim's work in which conjectures are
underpinned by perturbative arguments which can be carried out
for the especially simple boundary condition $\partial_1\phi =0$.}.
If one merely asks for what boundary conditions is a field theory
classically integrable, it might be expected there would be a class
of boundary conditions introducing a set of boundary parameters
in addition to the full-line parameters of the theory. Indeed,
this is apparently the case for the sine-Gordon model where the
most general integrable boundary condition contains two
free parameters
\NRF\rMv\Mv\refs{\rGZa ,\rMv} \foot{For reasons concerning stability
these may not be chosen entirely arbitrarily
\NRF\rCDRa\CDRa
\NRF\rFSa\FSa\refs{\rCDRa ,\rFSa}.} and is of the form:
\eqn\sgboundary{{\partial\phi\over\partial x^1}={a\over\beta}\sin \beta
\left({\phi -\phi_0 \over 2}\right)\qquad \hbox{at}\ x^1=0,}
where $a$ and $\phi_0$ are arbitrary constants, and $\beta$
is the sine-Gordon coupling. However, and surprisingly,
no other affine Toda field theory permits a full set of parameters
(ie equal to the rank of the Lie algebra whose root data defines the
model) in the boundary
\NRF\rCDRSa\CDRSa
\NRF\rBCDRa\BCDRa
condition \refs{\rCDRSa ,\rCDRa ,\rBCDRa}.
In fact, although the models based on
non simply-laced algebras do allow some free parameters in the
boundary condition, surprisingly few of the models permit boundary conditions
continuously connected to the Neumann condition
\eqn\neumann{{\partial\phi\over \partial x^1}\biggm|_{x^1=0}=0.}
Those that do are $c_n^{(1)},\ a_{2n}^{(2)},\ a_{2n-1}^{(2)},
\ d_4^{(3)}$ and $e_6^{(2)}$.

It was supposed, in \refs{\rCDRSa -\rBCDRa}, that the boundary
condition contained no time
derivatives, but  this could be too restrictive\foot{We are obliged to
Nick Warner for reminding us of this.}:  if we suppose there is
no kinetic energy specifically associated with
the boundary it is nevertheless possible,
provided there is more than one scalar field, to envisage  a boundary
condition which is linear in time derivatives, in addition to being
a function of the fields. In other words,
\eqn\oddboundary{{\partial\phi_a\over \partial x^1}=
-b_{ab}{\partial\phi_b\over \partial x^0}-{\partial{\cal B}\over
\partial\phi_a},\qquad \hbox{at}\ x^1=0,}
is a possible boundary condition,
corresponding to an additional term in the Lagrangian of the form
\eqn\boundaryterm{-\delta (x^1)\left({\cal B}(\phi ) +
{1\over 2}\phi_a b_{ab}\partial_0\phi_b\right) ,}
 where $b_{ab}$ is an antisymmetric matrix.
Such a boundary condition might be considered as the coupling of
a (constant) background antisymmetric tensor field. Locally, this
would be of the form
$$\partial_\mu\phi_a F^{\mu\nu}_{ab}\partial_\nu\phi_b ,$$
a total derivative if each component of
$F^{\mu\nu}$ satisfies free Maxwell equations.
On integration, it would lead to the boundary term above with
$b=F^{01}$. Boundary conditions of this general
type have been considered recently for free fields by Yegulalp
\NRF\rYa\Ya\refs{\rYa}. The two quantities
$b$ and ${\cal B}$ are to be determined by the requirements
of integrability. It will be seen below that \oddboundary\ with
$b\ne 0$ is
a rare possibility and is even more restrictive than the situation
with $b=0$. It will be seen there are no free parameters at
all for these cases. Nevertheless, because of the lack of
time-reversal invariance \oddboundary\ can provide examples
of classical reflection factors which differ between particle
and anti-particle;
such possibilities have been suggested previously by Sasaki on the basis
of the reflection bootstrap equations \refs{\rSk}.

In the case $b=0$, it was found
that the possible boundary conditions for the
$a_n^{(1)}, d_n^{(1)}$ and $e_n^{(1)}$ affine Toda
theories are highly constrained by the requirement that
there should be
conserved modifications of higher spin charges in the presence
of
the boundary. Effectively, in those cases, there is only a discrete ambiguity
and the possible boundary conditions are summarised by adding a term to the
action\foot{The notation and conventions for affine Toda field theory
are those of
\NRF\rBCDSc\BCDSc\refs{\rBCDSc}} of the form
\eqn\baction{{\cal L}_{\rm boundary}=-\delta (x^1) {\cal B}(\phi ),}
where
\eqn\allboundary{{\cal B}={m\over \beta^2}\sum_0^rA_ie^{{\beta\over 2}
\alpha_i\cdot\phi},}
and the coefficients $A_i,\ i=0,\dots ,r$ are a set of real numbers
with
\eqn\gboundary{\hbox{\bf either}\ |A_i|=2\sqrt{n_i}, \ \hbox{for}
\ i=0,\dots ,r\
\hbox{\bf or}\ A_i=0\ \hbox{for}\ i=0,\dots ,r\ .}
This result is also obtained by generalising the Lax pair idea to
include the boundary condition at $x^1=0$. Once the Lax pair is
available, all the other cases can be investigated and are listed in
\refs{\rBCDRa}.

To analyse the situation with $b\ne 0$, it is possible to proceed
in two directions. Firstly, it is not difficult to repeat, case by case,
the arguments of
\refs{\rCDRSa ,\rCDRa}, construct  conserved charges on the half-line using
low-spin conserved charges defined for the whole line, and find
constraints on the matrix $b$ and the boundary potential ${\cal B}$.
Alternatively, the Lax pair approach, mentioned earlier, can be adapted
to the present situation and used as a tool to determine the possible
choices of $b,\ {\cal B}$. In fact, in the case $b\ne 0$, the constraints
are sufficiently severe that the latter approach turns out to
provide, conveniently, a complete description of the classical problem.

\newsec{Boundary Lax pair}

The standard Lax pair for  affine Toda theory
\NRF\rMOPa{\MOPa\semi\Wa\semi\OTa}\refs{\rMOPa}
can be written
in the form
\eqn\laxfull{\eqalign{&a_0=H\cdot\partial_1\phi /2+\sum_0^r
\sqrt{m_i}(\lambda E_{\alpha_i}-1/\lambda \ E_{-\alpha_i}) e^{\alpha_i\cdot\phi
/2}\cr
&a_1=H\cdot\partial_0\phi /2+\sum_0^r
\sqrt{m_i}(\lambda E_{\alpha_i}+1/\lambda \ E_{-\alpha_i}) e^{\alpha_i\cdot\phi
/2},\cr}}
where $H_a, E_{\alpha_i}$ and $E_{-\alpha_i}$ are the Cartan subalgebra
and the generators
corresponding to the simple roots, respectively, of a simple Lie algebra of
rank $r$. Included in
the set of \lq simple' roots is the extra (affine) root, denoted $\alpha_0$,
which
satisfies
$$\sum_0^r\ n_i \alpha_i=0 \qquad n_0=1.$$
The coefficients $m_i$ are related to the $n_i$ by $m_i=n_i \alpha_i^2/8$.
The conjugation properties of the generators are chosen so that
\eqn\conj{a_1^\dagger  (x,\lambda )=a_1  (x,1/\lambda )
\qquad a_0^\dagger  (x,\lambda )
=a_0 (x,-1/\lambda ).}
Using the Lie algebra relations
$$[H, E_{\pm\alpha_i}]=\pm\, \alpha_i\, E_{\pm \alpha_i}\qquad
[E_{\alpha_i},E_{-\alpha_i}]=
2\alpha_i\cdot H/(\alpha_i^2),$$
the zero curvature condition for \laxfull\
$$f_{01}=\partial_0a_1-\partial_1a_0 +[a_0,a_1]=0$$
leads to the affine Toda field equations:
\eqn\todafull{\partial^2\phi =-\sum_0^r n_i \alpha_i e^{\alpha_i\cdot\phi}.}

To construct a modified Lax pair including the boundary condition
derived from \baction ,
it was found in \refs{\rBCDRa} to be convenient to
consider an additional  special point $x^1=b\ (>a)$ and two overlapping
regions $R_-:\ x^1\le (a+b+\epsilon )/2;\ $ and $R_+:\ x^1\ge (a+b-\epsilon
)/2$.
The second region will be regarded as a reflection of the first,
in the sense that if $x^1\in R_+$, then
\eqn\reflectphi{\phi (x^1)\equiv\phi (a+b-x^1).}
The regions overlap in a small interval surrounding the midpoint of $[a,b]$.
Then, in the two regions define:
\eqn\newlax{\eqalign{&R_-:\qquad \widehat a_0=a_0 -{1\over 2}\theta (x^1-a)
\left(\partial_1\phi +
{\partial\B\over\partial\phi}\right)\cdot H \qquad
\widehat a_1=\theta (a-x^1)a_1\cr
&R_+:\qquad \widehat a_0=a_0 -{1\over 2}\theta (b-x^1)
\left(\partial_1\phi -
{\partial\B\over\partial\phi}\right)\cdot H \qquad
\widehat a_1=\theta (x^1-b)a_1.\cr}}
Then, it is clear that in the region $x^1<a$ the Lax pair \newlax\ is
the same as the old but, at $x^1=a$ the derivative of
the $\theta$ function in the zero curvature condition enforces the boundary
condition
\eqn\boundary{{\partial\phi\over\partial x^1}=-{\partial{\cal B}\over \partial
\phi}, \qquad x^1=a .}
Similar statements hold for $x^1\ge b$
except that the
boundary condition at $x^1=b$ is slightly different  in order to
accommodate the reflection condition \reflectphi .

On the other hand, for $x^1\in R_-$ and $x^1>a$, $\widehat a_1$ vanishes
and therefore the zero curvature condition merely implies $\widehat a_0$
is independent of $x^1$. In turn, this fact implies  $\phi$ is
independent of $x^1$ in this region. Similar remarks apply to the region
$x^1\in R_+$ and $x^1<b$. Hence, taking into account the reflection principle
\reflectphi , $\phi$ is independent of $x^1$ throughout the interval $[a,b]$,
and equal to its value at $a$ or $b$. For general boundary conditions, a glance
at
\newlax\ reveals that the gauge potential $\widehat a_0$ is different in the
two
regions $R_\pm$. However, to maintain the zero curvature condition over the
whole
line the values of $\widehat a_0$ must be related by a gauge transformation
on the overlap. Since $\widehat a_0$ is in fact independent of $x^1\in [a,b]$
on both patches, albeit  with a different value on each patch,
the zero curvature condition effectively requires the existence of
a gauge transformation $\K$ with the property:
\eqn\Kdef{\partial_0 \K =\K\, \widehat a_0(x^0,b) -\widehat a_0(x^0,a)\, \K .}
The group element $\K$ lies in the group $G$ with Lie algebra $g$, the
Lie algebra whose roots define the  affine Toda theory.

The conserved quantities on the half-line ($x\le a$) are
determined via a generating function $\widehat Q(\lambda )$ given by the
expression
\eqn\Qalt{\widehat Q(\lambda )={\rm tr}\left( U(-\infty ,a;\lambda )\
\K \ U^\dagger(-\infty, a ;1/\lambda )\right),}
where $U(x_1,x_2;\lambda )$ is defined by the path-ordered exponential:
\eqn\pathexp{U(x_1,x_2;\lambda )={\rm P}\exp \int_{x_1}^{x_2}\, a_1 dx^1.}

Assuming $\K$ is independent of both $x^0$ and the fields $\phi$, or
their derivatives,
 eq\Kdef\ reads,
\eqn\Kdefa{{1\over 2}\left[\K (\lambda ),\,
{\partial\B\over\partial\phi}\cdot H\right]_+=-\,\left[\K (\lambda )
,\, \sum_0^r
\sqrt{m_i}(\lambda E_{\alpha_i}-1/\lambda \ E_{-\alpha_i})
e^{\alpha_i\cdot\phi /2}\right]_-,}
where the field dependent quantities are evaluated at the boundary $x^1=a$.
Eq\Kdefa\ is rather restrictive, since the boundary term $\B$ does not
depend on the spectral parameter $\lambda$, and leads to the results
concerning the boundary potential claimed in eqs\allboundary\ and
\gboundary , and given in detail elsewhere.

\newsec{Modified boundary Lax pair}

It is useful to note  the Lax pair \laxfull\ is not unique.
In particular, since the curvature $f_{01}$ lies in the Cartan
sub-algebra spanned by the generators $H_a$, the affine Toda equations of
motion will still be obtained if the \lq gauge' fields $a_0,a_1$ are gauge
transformed using any group element of the form
$$g=e^{i\theta\cdot H}.$$ However, this is not true of the
boundary condition, coded via \newlax\ in the
modified fields $\widehat a_0,\widehat a_1$. Indeed, if the condition
\boundary\  is replaced by the condition \oddboundary , or
its reflected version, in
the definitions of $\widehat a_0$ in the two overlapping
regions, then the additional terms proportional to
$H_ab_{ab}\partial_0\phi_b$ can be removed by making a gauge
transformation based on the group elements:
\eqn\boundarygauge{g_\pm=e^{\pm H\cdot b\phi}}
in the regions $R_\pm$, respectively. After performing these
gauge transformations, in the overlap region $a<x^1<b$ one finds:
\eqn\newao{\eqalign{&\widehat a_0^\prime = -{1\over 2}\,
{\partial{\cal B}\over
\partial\phi}\cdot H +\sum_0^r \sqrt{m_i}\left[\lambda
e^{\alpha_i(1+b)\cdot\phi /2}-{1\over \lambda}e^{\alpha_i(1-b)\cdot\phi /2}
E_{-\alpha_i}\right]\quad x^1\in R_-\cr
&\widehat a_0^\prime = \phantom{-}{1\over 2}\,
{\partial{\cal B}\over
\partial\phi}\cdot H +\sum_0^r \sqrt{m_i}\left[\lambda
e^{\alpha_i(1-b)\cdot\phi /2}-{1\over \lambda}e^{\alpha_i(1+b)\cdot\phi /2}
E_{-\alpha_i}\right]\quad x^1\in R_+.\cr}}
These two, modified, gauge fields must then be related by $\K$ according
to eq\Kdef .

If it is further assumed $\K$ does not depend upon $x^0$ or $\phi$, then
\Kdefa\ is modified to read:
\eqn\Kdefb{\eqalign{{1\over 2}\left[\K,\,
{\partial\B\over\partial\phi}\cdot H\right]_+=-\,\sum_0^r\sqrt{m_i}
&\left[ e^{\alpha_i(1-b)\cdot\phi /2} \left(\lambda\K E_{\alpha_i}
+{1\over \lambda}E_{-\alpha_i}\K\right)\right.\cr
&\left. -e^{\alpha_i(1+b)\cdot\phi /2}
\left({1\over\lambda}\K E_{-\alpha_i} +\lambda E_{\alpha_i}
\K\right)\right] . \cr}}
For the specific cases of interest, $\phi$ in \Kdefb\ refers to the values of
the field at $x^1=0$.

Next, suppose $\K (0)$ exists (at least after multiplying $\K$ by a suitable
power of $\lambda$). Then, \Kdefb\ reduces to
\eqn\Kzero{\sum_0^r\sqrt{m_i}\left[ e^{\alpha_i(1+b)\cdot\phi /2}\K (0)
E_{-\alpha_i}
-e^{\alpha_i(1-b)\cdot\phi /2}E_{-\alpha_i}\K (0)\right]=0,}
and, since $\K$ is independent of the field value at the boundary, this
in turn implies two conditions for each of $i=0,\dots ,r$:
\eqn\bcondition{\alpha_i(1-b)=\alpha_{\pi (i)} (1+b)}
where $\pi$ is a permutation of $0,\dots ,r$ and,
\eqn\Kzerocondition{\K (0)\sqrt{m_i}\, E_{-\alpha_i}\K^{-1}(0)=
\sqrt{m_{\pi (i)}}\,
E_{-\alpha_{\pi (i)}}.}
The first of these conditions, \bcondition , implies  $\pi$ is
an automorphism of
the extended
Dynkin diagram whose root system defines the affine Toda theory under
discussion (therefore $\sqrt{m_i}=\sqrt{m_{\pi (i)}}$); the second,
\Kzerocondition ,
requires $\pi$ to be an inner
automorphism. In other words, $\pi$ is a symmetry of the extended Dynkin
diagram which is not also a symmetry of the Dynkin diagram itself---the group
of such symmetries being isomorphic to the centre of the Lie group.
{}From these observations, it is already very clear the conditions
\oddboundary\ are only rarely compatible with integrability. Indeed,
the field theories which might allow integrable boundary
conditions of this type may only be chosen from
$a_r^{(1)},d_r^{(1)},e_6^{(1)},e_7^{(1)},b_r^{(1)},c_r^{(1)},
a_{2r-1}^{(2)},d_{r+1}^{(2)}$.

However, examining \bcondition\ carefully reveals that only
odd order automorphisms are admissable. To see this, suppose $\pi$
has order $p$
and consider
\eqn\brels{\alpha_i(1-b)=\alpha_{\pi (i)} (1+b),\ \alpha_{\pi (i)}(1-b)=
\alpha_{\pi^2 (i)} (1+b),\  \dots\
\alpha_{\pi^{p-1} (i)}(1-b)=\alpha_i (1+b),}
and take the alternating sum to find
\eqn\podd{\alpha_i b=\alpha_{\pi (i)}-\alpha_{\pi^2 (i)}+\
\dots\ -\alpha_{\pi^{p-1}(i)},}
if $p$ is odd and,
\eqn\peven{\alpha_{\pi (i)}-\alpha_{\pi^2 (i)}+\ \dots\ +
\alpha_{\pi^{p-1}(i)}=0,}
when $p$ is even. Adding \peven\  to a similar equation with $i$
replaced by $\pi (i)$
immediately implies
$$\alpha_{\pi (i)}=-\alpha_i,$$
clearly impossible for a set of simple roots. Given the relevant
automorphisms
have odd order, the set
of possible data is restricted to a choice from at most $a_r^{(1)}$
and $e_6^{(1)}$.

Since $\K (0)$ represents an inner automorphism of the Lie algebra
of a compact Lie group,
choose it to be unitary. Then, using \Kzerocondition\ it follows that
\eqn\KzeroH{\K (0)\alpha_i\cdot \K^{-1}(0)= \alpha_{\pi (i)}\cdot H .}

Setting
\eqn\Kfirst{\K (\lambda )= (1+k_1\lambda +k_2\lambda^2+
O(\lambda^3))k_0,}
and examining the order $\lambda$ terms in \Kdefb , leads to an
equation determining
both $k_1$ and the boundary potential ${\cal B}$:
\eqn\Konecondition{\eqalign{{1\over 2}\, {\partial{\cal B}\over
\partial\phi}\cdot
\left( H+\right. & \left. k_0Hk_0^{-1}\right)\cr
&=\sum_0^r\sqrt{m_i}\left[ e^{\alpha_i(1+b)\cdot\phi /2}\, k_1\,
k_0E_{-\alpha_i}k_0^{-1}-
 e^{\alpha_i(1-b)\cdot\phi /2} E_{-\alpha_i}\, k_1\right]\cr
&=\sum_0^r\sqrt{m_i} e^{\alpha_i(1-b)\cdot\phi /2} [k_1,\,
E_{-\alpha_i}],\cr}}
the second step following immediately from \bcondition\ and
\Kzerocondition .
Clearly, considering the grading of the Lie algebra generators on
the two sides
of \Konecondition , bearing in mind \KzeroH , $k_1$ must be a
linear combination of
the positive simple root step operators and the step operator
corresponding to the
lowest root $\alpha_0$:
\eqn\Kone{k_1 =\sum_0^r (A_i/\sqrt{m_i}) E_{\alpha_i},}
where the $A_i$ are a set of constants.
Therefore, using \KzeroH\ and \bcondition , \Konecondition\  reduces
to an equation
constraining ${\cal B}$
\eqn\Bcondition{{1\over 2}\, {\partial{\cal B}\over \partial\phi}
\cdot\left(
H+(1+b)(1-b)^{-1}H\right)
=\sum_0^r\, \alpha_i\cdot H e^{\alpha_i(1-b)\cdot\phi /2}.}
Matching the coefficients of the independent elements of the Cartan
subalgebra, and multiplying
through by $1-b$ yields
$${\partial{\cal B}\over \partial\phi}=\sum_0^r A_i\alpha_i (1-b)
e^{\alpha_i(1-b)\cdot\phi /2},$$
which implies
\eqn\Bexpression{{\cal B}=2\, \sum_0^r A_i e^{\alpha_i(1-b)\cdot\phi /2}.}
Thus, provided the coefficients $A_i$ are not subsequently forced to vanish,
the boundary
potential again has the characteristic exponential form although the exponents
may contain vectors other than simple roots. If $b=0$, \Bexpression\ reduces to
the
results obtained before \refs{\rCDRa, \rCDRSa ,\rBCDRa}.

Using the expression for ${\cal B}$, and  assuming $\K$ is independent of
$\phi$, leads to a set of equations, one for each $i=0,1,\dots , r$,
from which any
further constraints on $\K (\lambda )$ will be derived:
\eqn\Kgeneral{\eqalign{{A_i\over 2}\, \left[\K (\lambda ),\,
\alpha_i(1-b)\cdot H\right]_+=
&-\lambda\, \K (\lambda )\,  E_{\alpha_i}+ {1\over \lambda}\, \K (\lambda )\,
E_{-\alpha_{\pi^{-1}(i)}}\cr &+\lambda\ E_{\alpha_{\pi^{-1}(i)}}\,\K (\lambda )
-
{1\over \lambda}\, E_{-\alpha_i}\,\K (\lambda ).\cr}}

To analyse further the  generic set of cases, $a_r^{(1)}$, it is convenient to
work in the fundamental representation and to introduce
the following pair of matrices $P$ and $Q$,
\eqn\defPQ{P=\pmatrix{0&1&0&0&\dots&0\cr
                                     0&0&1&0&\dots&0\cr
                                     0&0&0&1&\dots&0\cr
                                     0&0&0&0&\dots&1\cr
                                     1&0&0&0&\dots&0\cr}
              \quad Q=\hbox{Diag}(1,\omega ,\omega^2 ,\omega^3 ,
\dots ,\omega^r),
\quad \omega^{r+1}=1,}
which satisfy
\eqn\PQrelations{P^{r+1}=Q^{r+1}=1, \qquad PQ=\omega QP.}
In terms of these,
the generators corresponding to the simple roots are given by
\eqn\Egenerators{E_{\alpha_k}={1\over r+1}\, \sum_{s=0}^r \omega^{-ks}P
Q^s,\qquad k=0,1,\dots ,r,}
and the elements of the Cartan subalgebra are
\eqn\Hgenerators{\alpha_k\cdot H={1\over r+1}\, \sum_{s=0}^r
(\omega ^s -1)\omega^{-ks} Q^s, \qquad k=0,1,\dots ,r.}
Using \PQrelations\ and \Egenerators , it is easy to check $P$ implements
an elementary cyclic permutation of the generators corresponding
to the simple roots:
\eqn\Perm{\eqalign{P^{-1}E_{\alpha_k}P&={1\over r+1}\, \sum_{s=0}^r
\omega^{-ks}P
 \omega^{-s}Q^s=E_{\alpha_{k+1}}\cr
P^{-1}\alpha_k\cdot HP&=\alpha_{k+1}\cdot H ,\cr}}
and, therefore, the set $P^s,\ s=0,1,2,\dots ,r$ are the elements
of the $Z_{r+1}$ group of symmetries of the $a_r^{(1)}$ extended
Dynkin-Ka\v{c} diagram.

Suppose the permutation $\pi$ in eq\Kgeneral\ is represented by $P^L$
then, using \Egenerators\ and \Hgenerators , eq\Kgeneral\ may be rewritten
usefully as follows:
\eqn\Kgenerala{\eqalign{\sum_s\omega^{-ks}&\left({A_k\over 2} (\omega^s-1)
(1-\omega^{Ls}+\omega^{2Ls}-\dots +\omega^{(p-1)Ls})
\left[\K ,Q^s\right]_+\right.\cr &\left. +\lambda \K P Q^s
-{1\over \lambda} \omega^{-Ls}\K Q^s P^{-1}-
\lambda \omega^{-Ls}P Q^s \K +{1\over\lambda}Q^s P^{-1}\K\right)=0.\cr}}
Given the form of eq\Kgenerala , it seems natural to take, as
an ansatz for $\K$,
\eqn\Kansatz{\K (\lambda )=\sum_tk_t(\lambda )P^t,} and to suppose
the coefficients of $\omega^{-ks}Q^sP^t$ vanish in eq\Kgenerala\ for
each choice of $k,s$ and $t$. In other words,
\eqn\Kgeneralb{\eqalign{{A_k\over 2}(\omega^s -&1)(1+\omega^{ts})
(1-\omega^{Ls}+\omega^{2Ls}-\dots +\omega^{(p-1)Ls})k_t \cr
&+\lambda k_{t-1} \omega^{ts}-\lambda k_{t-1}\omega^{(1-L)s}-
{1\over\lambda}k_{t+1}\omega^{(t+1-L)s}+
{1\over\lambda}k_{t+1}=0 ,\cr}}
where the coefficients $k_t$ depend upon $t $ and $\lambda$ but do not
depend upon $s$ or $k$. Indeed, the only dependence on the label
$k$ occurs in the coefficients $A_k$ which must therefore be equal to
each other (they may all be zero). Apart from the latter remark, there
are several cases.
\bigskip
\noindent{\bf I: $A_k=0\ k=0,1,\dots ,r$}

In this case, \Kgeneralb\  reduces to
$$k_{t+1}(1-\omega^{(t-L+1)s})=\lambda^2 k_{t-1}(\omega^{(1-L)s}-
\omega^{ts}),$$
which has a solution provided $L=-1$ (and, therefore, $r$ must be even),
and all but two of the coefficients
($k_1$ and $k_r$) are zero. Ie $\K$ is given by
\eqn\Kone{\K (\lambda )=\lambda P -{1\over\lambda}P^{-1},}
and $b$ is given by \podd , with $\pi =P^{-1}$.

\bigskip

\noindent{\bf II: $A_0=A_1=A_2=\dots =A_r=A$}

In this case, \Kgeneralb\ may be reorganised by multiplying through
by $1+\omega^{Ls}$, to obtain:
$$\eqalign{&A (\omega^s -1)(1+\omega^{ts})k_t \cr
&+(1+\omega^{Ls})\left(\lambda
k_{t-1} \omega^{ts}-\lambda k_{t-1}\omega^{(1-L)s}-{1\over\lambda}
k_{t+1}\omega^{(t+1-L)s}+
{1\over\lambda}k_{t+1}\right)=0, \cr}$$
which is solved (assuming none of the coefficients vanish) provided
$$A k_t=\lambda k_{t-1} ={1\over\lambda}k_{t+1}, \qquad \omega^{2Ls}=
\omega^s ,$$
in turn implying
\eqn\Aconditions{A^2=1, \qquad L={r+2\over 2}.}
Again, $r$ must be even, and $\K$ is given by
\eqn\Ktwo{\K (\lambda )=\sum_{-r/2}^{r/2} (A\lambda P)^t.}
\bigskip

\noindent{\bf III: $r=5$}

Special cases may occur in II if some of the coefficients $k_t$
in fact vanish. However, direct inspection reveals there is precisely
one such case for which
\eqn\specialcase{r=5, \qquad L=-2, \qquad A^2=1, }
and
\eqn\Kthree{ \K(\lambda )
={1\over\lambda^2}P^{-2}-{A\over \lambda}P^{-1}+A\lambda P-\lambda^2 P^2.}
For this, the permutation $\pi$ is of order three.
\bigskip
To analyse the remaining case, $e_6^{(1)}$, return to the perturbative
expansion of $\K$, eq\Kfirst , and attempt to determine the
term at order $\lambda^2$, given \Kone\  and \Bexpression .
After some manipulation, one obtains the set of equations:
\eqn\Ktwocondition{\eqalign{\left[ k_2,E_{-\alpha_i}\right] =
E_{\alpha_{\pi (i)}}-&E_{\alpha_{\pi^{p-1} (i)}}\cr &-{A_i\over 2}
\left(\alpha_{\pi (i)}-\alpha_{\pi^2 (i)}+\dots -\alpha_{\pi^{p-1} (i)}\right)
\cdot \sum_j A_j\alpha_j E_{\alpha_{j}},\cr}}
for each of $i=0,1, \dots , r$. Clearly, $k_2$ must be a linear combination of
generators corresponding to level two roots. The permutation $\pi$ is the
threefold symmetry of the extended $e_6^{(1)}$ diagram whose orbits consist
of the three outer roots (labelled 0,1,2) the three inner roots (labelled
3,4,5) and the centre root (labelled 6), taken clockwise with the pairs (0,3),
(1,4) and (2,5) lying on the three legs, respectively.
Examining, eqs\Ktwocondition\ for $i=0,1,2$ leads immediately to the
conclusion $A_3=A_4=A_5=0$. However, the equations corresponding to
i=3,4,5 are then inconsistent; for example, when $i=3$
$$\left[ k_2, E_{-\alpha_3}\right] =E_{\alpha_4}-E_{\alpha_5}$$
which may never be satisfied since $\alpha_4+\alpha_3$ and
$\alpha_5+\alpha_3$ are not roots. Hence, for $e_6^{(1)}$ the hypothesis
concerning the existence of $\K (0)$ is false, and $\K$ cannot
exist.

\newsec{Discussion}

The linearised version of the field equations and the
boundary conditions may be examined, noting that in all the
allowable cases
I, II and III, $\phi^{(0)} =0$ is a possible classical solution.

First, note that the matrix $b$ and the mass matrix of the affine Toda
field theory commute. The mass matrix $M$ is defined by
\eqn\massmatrix{M^2=\sum_0^rn_i\alpha_i\otimes\alpha_i,}
and therefore, using the antisymmetry of $b$, the commutator
\eqn\bMcomm{\left[ M^2,b\right] =\sum_0^rn_i\left(\alpha_i\otimes\alpha_i b
+\alpha_ib\otimes\alpha_i\right) ,}
can be evaluated using \podd . Rearranging the sums using the permutation
$\pi$, and
recalling $\pi$ has odd order, leads to the terms on the right hand side of
\bMcomm\  cancelling pairwise to zero. Actually, this fact was to be expected
since the mass-matrix is invariant under permutations of the roots
corresponding to symmetries of the Dynkin, or extended Dynkin-Ka\v c, diagram
and $b$ is directly related to such a symmetry.

Second, to express $b$  in terms of $\pi$,
consider the
latter as
a linear mapping of the roots:
\eqn\pimap{\alpha_{\pi (i)}=\widehat\pi \alpha_i,}
and use eq\brels , to deduce:
\eqn\bpi{b={\widehat\pi -1\over \widehat\pi +1},}
from which, knowing the eigenvalues of $\widehat\pi$,
the eigenvalues of $b$ may be read off. In every case, $\widehat\pi$ is a
power of the
permutation matrix which, acting on the roots, has  eigenvalues
which are the $(r+1)$st roots of unity, except $1$ itself.
Hence the eigenvalues of $b$, $b_s,\  s=1,2,\dots ,r,$ are
\eqn\beigenvalues{b_s=i\tan\left({\pi L s\over r+1}\right)=-b_{r+1-s}.}

Case I is the simplest to treat because there is no boundary potential.
There is a scattering solution to the linearised problem, of the form:
\eqn\solution{\phi^{(1)} =\epsilon \sum_s \rho_s e^{-i\omega_s x^0}\left( R_s
e^{-ik_s x^1} +e^{ik_s x^1} \right) \qquad x^1<0,}
where
$\rho_s$ are the common eigenvectors of the mass$^2$ matrix and the
boundary matrix $b$:
$$M^2\rho_s=m_s^2 \rho_s \qquad b\rho_s =b_s \rho_s,$$
and
$$\omega_s=m_s\cosh\theta\qquad k_s=m_s\sinh\theta .$$
Using \beigenvalues , with $L=-1$, and the boundary condition
\oddboundary , yields an expression for the reflection factors $R_s$:
\eqn\Ireflection{R_s ={k_s-b_s\omega_s\over k_s+b_s\omega_s}=-{(s)
\over (r+1-s)},}
where the final step uses the notation
$$(x)={\sinh \left( {\theta\over 2}+{i\pi x\over 2(r+1)}\right)\over
\sinh \left( {\theta\over 2}-{i\pi x\over 2(r+1)}\right)}.$$

Notice, as a consequence of \beigenvalues , the reflection factors for a
particle and its conjugate are not equal; rather $R_{\bar s}=R_s^{-1}=
R_s(i\pi -\theta )$.
Clearly, a distinction between the two classical reflection factors was
to be expected since the boundary condition is not time-reversal invariant.
Some time ago, Sasaki \refs{\rSk} discovered asymmetric solutions to the
reflection
bootstrap equations but without noting examples of boundary conditions
which might be responsible for them. Notice also, these reflection
factors satisfy the classical limit of the reflection bootstrap equations
for the  $a_r^{(1)}$ theories. In other words, if the three particles
$r,s,t$ couple in the quantum field theory (in the sense that any of
them, say $r$, may be a bound state of the other two,  $s$ and $t$)
then, assuming factorisability, the reflection bootstrap equations
provide an expression for $K_r$ in terms of $K_s$ and $K_t$:
\eqn\bootstrap{K_r(\theta )= K_s(\theta -i\bar\theta^t_{sr})
K_t(\theta +i\bar\theta^s_{tr})
S_{st}(2\theta ).}
The classical limit of \bootstrap\ replaces $K$ by $R$ and $S$ by unity.
The notation, and the data concerning coupling angles necessary to verify
the assertion can be found in \refs{\rBCDSc}.

For cases II and III, the linearised boundary condition is more
complicated
and for each mass eigenvalue
takes the form \eqn\IIcondition{\partial_1\phi_s=-b_s\partial_0\phi_s -
{A\over
2}
(1-b_s^2)m_s^2
\phi_s\qquad x^1=0.}
For example, in case III, $b_s=-i\tan (\pi s/3)$ and
the linearised reflection factors are given by
\eqn\IIIcondition{R_s={\sinh \theta \cos^2 (\pi s/3) +i\cosh\theta\cos(\pi s/3)
\sin(\pi s/3)+iA\sin(\pi s/6)\over \sinh \theta \cos^2 (\pi s/3) -
i\cosh\theta\cos(\pi s/3)
\sin(\pi s/3)-iA\sin(\pi s/6)}.}
Curiously, taking $A=1$ for instance, one finds $R_1=(5)^2$ in the standard
notation but, the classical version of the reflection bootstrap equation
is not actually satisfied since $R_2\ne (4)^2$. In this instance,
it appears the solution to the linear problem is not the classical
limit of a quantum theory with a factorisable S-matrix and reflection
factors; perhaps the quantum field theory is not integrable
in these cases.
Similar remarks apply to reflection factors derived in case II.

It appears classical integrability in the presence of boundary
conditions is a rare phenomenon and it is curious that very few
of the known examples (apart from sinh/sine-Gordon) permit either a
continuous deformation away from the Neumann condition,
$\partial_1\phi\big|_{x^1=0}=0$, or a continuous deformation away from
the free or from the conformal Toda situation. It seems a
\lq quantisation' of the boundary parameters is largely inescapable.

\bigskip

\noindent{\bf Acknowledgements}
\bigskip
We would like to thank Nick Warner for a stimulating
conversation and
two of us (PB and RHR) are grateful to the Engineering and Physical
Sciences Research Council for post-doctoral support. We are also
grateful for partial support from the EC Commission via a Human Capital
and Mobility Grant, contract  number  ERBCHRXCT920069.

\bigskip

\noindent{\bf Appendix}
\bigskip
There is a direct argument implying that the classical reflection factors
should satisfy a bootstrap condition. Since the argument does not seem to
appear elsewhere in the literature, it will be included in outline
here for completeness.

Suppose $\phi^{(0)} =0$ is a solution to the full field equations, plus
a boundary condition at $x^1=0$. Then, assuming there is another solution which
may be considered to be small, it has an expansion (in terms of the coupling
constant if one prefers), of the type:
$$\phi =\phi^{(1)} +\phi^{(2)} + \dots \ .$$
The first two terms satisfy the equations:
\eqn\firsttwo{\eqalign{&(\partial^2 +M^2)_{ab}\phi^{(1)}_b =0 \cr
                       &(\partial^2 +M^2)_{ab}\phi^{(2)}_b
=-c^{abc}\phi^{(1)}_b
\phi^{(2)}_c,\cr}}
where $c^{abc}$ are the classical couplings to be found in \refs{\rBCDSc}.
With a boundary, the solutions sought are perturbations of the
solutions to the linear equation given in \solution . Thus,
\eqn\nextorder{\eqalign{\phi^{(2)}=\epsilon^2&\sum_{r,s,t}\rho_r \,
\widehat{c}_{rst}
\, e^{-i(\omega_s+\omega_t)x^0}\,  \times \cr
&\left(\, {1\over
(\omega_s+\omega_t)^2-(k_s+k_t)^2-m_r^2}\left[e^{i(k_s+k_t)x^1}
+R_s R_te^{-i(k_s+k_t)x^1}\right]\right.\cr
&+\left.{1\over
(\omega_s+\omega_t)^2-(k_s-k_t)^2-m_r^2}\left[R_te^{i(k_s-k_t)x^1}
+R_se^{-i(k_s-k_t)x^1}\right]\right) ,\cr}}
where
$$\widehat{c}_{rst}=c^{abc}\rho^a_r\rho^b_s\rho^c_t,$$
and it has been assumed the eigenvectors $\rho_s^a$ are normalised to
unity.

Clearly,  the first term on the right hand side of \nextorder\
has a pole when the momenta and energy of particles $s$ and $t$
happen to lie on the mass-shell of particle $r$, and the term exists in
the sum for a particular $s$ and $t$ with a classical coupling to $r$.
If this can happen, the term $\phi^{(2)}$ dominates and consistency
with the boundary conditions
would require the coefficient of the pole to agree with the leading order
reflection coefficient of particle $r$. In other words, one is led
to deduce a classical bootstrap condition
$$K_r(\theta )= K_s(\theta -i\bar\theta^t_{sr})
K_t(\theta +i\bar\theta^s_{tr}),$$
reminiscent of the bootstrap property of the soliton solutions
in the complex affine Toda field theory
\NRF\rHa\Ha\refs{\rHa}. Apparently, the classical bootstrap
property depends only on the field equations in the region $x<0$.
The difficulty with this argument rests with the boundary condition.
The first order approximation has been designed to satisfy the
boundary condition at $x=0$ but there is no guarantee that the next
order term, determined by eq\nextorder , will do so. In general, it
may not and
an extra term of order $\epsilon^2$, satisfying the homogeneous
equation,  must be added to \nextorder\ to maintain the boundary
condition.

Returning to the two examples given in section 4, the first, for
which the reflection coefficient is given by \Ireflection , leads
to a second order term \nextorder\ which does in fact satisfy
the boundary condition automatically. Although it has not been
checked beyond the second order, one suspects the (minimal) perturbative
solution satisfies the boundary condition order-by-order in this case.
On the other hand, the second example, for which the reflection coefficient
is given by \IIIcondition , does not lead to a perturbative solution
satisfying the boundary condition without the explicit addition of extra pieces
at each order.

\listrefs

\end